\pdfoutput=1
\documentclass[aip, jap, reprint, amsmath, amssymb]{revtex4-1}
\usepackage{graphicx}
\usepackage{dcolumn}
\usepackage{bm}
\usepackage[utf8]{inputenc}
\usepackage[T1]{fontenc}
\usepackage{mathptmx}
\usepackage[pagewise]{lineno}

\begin{document}

%--------------------------------------------------------------------------------------------------------------------------------------------------------------------------------
\title{Identification of point defects on Co-Ni co-doping in SnO$ _{2}$ nanocrystals and their effect on the structural and optical properties}

%--------------------------------------------------------------------------------------------------------------------------------------------------------------------------------
\author{S. Roy}
 \affiliation{Department of Physics, Institute of Science, Banaras Hindu University, Varanasi -- 221005, India.}
\author{Brijmohan Prajapati}
  \affiliation{Department of Physics, Institute of Science, Banaras Hindu University, Varanasi -- 221005, India.}
\author{A. Singh}
 \altaffiliation[Presently at: ] {Department of Condensed Matter Physics and Materials Science, Tata Institute of Fundamental Research, Mumbai -- 400005, India.}
 \affiliation{Central Instrument Facility, Indian Institute of Technology (Banaras Hindu University), Varanasi - 221005, India.}
\author{Amish G. Joshi}
 \affiliation{CSIR-National Physical Laboratory, Dr. K. S. Krishnan Road, New Delhi -- 110012, India.}
\author{S. Chatterjee}
 \affiliation{Department of Physics, Indian Institute of Technology (Banaras Hindu University), Varanasi -- 221005, India.}
\author{Anup K. Ghosh}
    \email{akghosh@bhu.ac.in}
      \affiliation{Department of Physics, Institute of Science, Banaras Hindu University, Varanasi -- 221005, India.}
      

%-------------------------------------------------------------------------------------------------------------------------------------------------------------------------------------
%\date{\today}
             
%--------------------------------------------------------------------------------------------------------------------------------------------------------------------------------------
\begin{abstract}
Sn$ _{0.97-y} $Co$ _{0.03} $Ni$ _{y} $O$ _{2} $ (0 $  \leq$ \textit{y} $  \leq$ 0.04) nanocrystals, with the average crystallite size in the range 7.3 nm (for, \textit{y} = 0.00) -- 5.6 nm (for, \textit{y} = 0.04), have been synthesized using pH-controlled chemical co-precipitation technique. All the non-stoichiometric and stoichiometric point defects arising in the nanocrystals on co-doping have been identified and their effect on structural and optical properties of the nanocrystals have been extensively studied. It has been observed, using X-ray photoelectron spectroscopy (XPS), that on increasing the Ni co-doping concentration (\textit{y}), the non-stoichiometric Sn defect, Sn$_{\text{Sn}}^{''}$, increases in compensation of the existing defect Sn$_{i}^{\centerdot\centerdot\centerdot\centerdot}$ for \textit{y} = 0.00 nanocrystals. High resolution transmission electron microscopy (HR-TEM) also confirms the existence of Sn$_{\text{Sn}}^{''}$. Regarding the stoichiometric Frenkel defect, XPS  results have indicated that the concentration of \textit{V}$ _{\text{O}}$ and O$ _{i}$, manifested in the form of dangling bond related surface defect states, increases with increase in \textit{y}. Temperature dependent magnetisation measurement of the nanocrystals confirm the charge state of \textit{V}$ _{\text{O}}$. The point defects have been found to affect the structural properties in a way that the distortion in octahedral geometry of complete Sn -- O octahedron effectively reduces whereas the distortion in the trigonal planar coordination geometry of oxygen increases.  A direct effect of the O related Frenkel defect has been observed on the blue luminescence of the nanocrystals such that the spectral contribution of blue luminescence in the total emission intensity increases by $ \approx $ 72\% for \textit{y} = 0.04 as compared to \textit{y} = 0.00. 
\end{abstract}

%---------------------------------------------------------------------------------------------------------------------------------------------------------------------------------------
\maketitle

\section{Introduction} 
Technologically important wide band gap oxide semiconductors have attracted immense interest because of their diverse physio-chemical properties, and an ease of fabrication as compared to their traditional non-oxide counterparts \cite{H_Peng, T_Fukumura}. Among them, SnO$ _{2} $, because of its chemical stability, a large optical band gap ($ \approx $ 3.6 eV) and n-type conductivity with a high carrier density ($ \sim $ 10$ ^{20} $ cm$ ^{-3} $), has been applied in a large range of device applications \cite{J_Jiang, Y_Feng, Meifang}. 

SnO$ _{2} $ is a IV-VI semiconductor. It crystallises in tetragonal rutile structure with space group P4$_{2}$/\textit{mnm}, where each Sn(IV) ion is placed at the centre of a slightly distorted oxygen octahedron and each O(II) ion is having trigonal planar coordination geometry of Sn(IV) ions \cite{Godinho}. The unique coexistence of optical transparency ($ \sim $ 97\% of visible light) with high electrical conductivity in SnO$ _{2} $ is related to its inherent non-stoichiometry. Based on the experimental results it has been concluded that oxygen-related intrinsic point defects is the cause of this non-stoichiometry \cite{Samson}. Many contradictory findings on the exact nature and origin of these defects have been reported so far. In one of the reports \cite{Kilic}, the oxygen deficiency in SnO$ _{2} $ was attributed to Sn interstitials (Sn$_{i}$ \cite{Kroger}), which lowers the formation energy of oxygen vacancies (V$ _{\text{O}} $) in the lattice. It was also concluded that Sn$_{i}$ and \textit{V}$ _{\text{O}} $ coexist because of a strong mutual attraction ($ \approx $ 3.2eV) between them. In another report, \textit{V}$ _{\text{O}} $ was found to have lower formation energy than Sn$_{i}$ and the cluster defect of Sn$_{\text{Sn}}^{''}$ + V$ _{\text{O}}^{\centerdot\centerdot}$ was calculated to be dominating because of their lowest formation energy (in O-rich conditions) and cause the overall oxygen deficiency in the lattice \cite{Godinho}. In yet another report, the Sn$_{i}$ and \textit{V}$ _{\text{O}} $ were discarded as the cause of electrical conductivity of SnO$ _{2} $ and was attributed to incorporation of hydrogen impurities at interstitial sites or at lattice oxygen sites \cite{AKSingh}. In a recent report, using a hybrid quantum mechanical/molecular mechanical (QM/MM) embedded cluster approach, it has again been shown that oxygen vacancies, forming deep states in SnO$ _{2} $ contribute significantly to its intrinsic n-type conductivity \cite{Buckeridge}. However, despite all such reports based on first-principles calculations, proper conclusive experimental findings on the nature and composition of intrinsic defects of SnO$ _{2} $ are scarcely available. Recently, it has been shown, based on experimental results, that for Sn$  _{2}$(Nb$  _{2-x}$Ta$  _{x}$)O$  _{7}$ (\textit{x} = 0 -- 2) -- a multinary tin oxide-based transparent conducting oxide, controlling the lattice defects can help in tuning the optical band gap and carrier density \cite{Kikuchi}. Thus, it can be assumed for SnO$ _{2} $ that its intrinsic point defects can also have a profound effect on the structure and optical properties, a detailed study of which is lacking.

SnO$ _{2} $ in pristine form is rarely used and dopants, such as, Co \cite{Ogale,Kuldeep,Roy}, Ni \cite{Ateeq,Srinivas,Archer}, Cu \cite{Meifang}, Fe \cite{Kuldeep}, Mn \cite{Radovanovic2}, \textit{etc}. have been employed at the cationic site to tailor the physical and chemical properties.  However, due to charge imbalance and crystal radius mismatch between the dopant and host, the doping at high concentrations get limited \cite{Roy}. As such, co-doping has been proposed as an effective means to increase the dopant solubility as well as increasing the carrier mobility in the host matrix \cite{Zhang_Jing}. Excellent magnetic properties have been observed in cobalt doped SnO$ _{2} $ nanostructure \cite{Ogale}. But higher concentration of Co doping is not achieved \cite{Roy} because of the large ionic radius mismatch between Co ($ \approx $ 0.545 \AA) and Sn ($ \approx $ 0.69 \AA) \cite{Shannon}. Because ionic radius of Ni ($ \approx $ 0.56 \AA) is somewhat closer to Sn ($ \approx $ 0.69 \AA) \cite{Shannon}, it can act as better substitute \cite{Kiyoshi_Fe-Ni}. In fact, Ni doping in SnO$ _{2} $ nanocrystals has also shown some excellent optical and magnetic properties \cite{Ateeq}. However, although it can be expected that these properties are largely influenced by the point defects, only few experimental reports are available concerning the identification and effect of point defects on doping, for example, the identification and effect of point defects on structural and optical properties have been extensively studied for Co doped SnO$ _{2} $ nanocrystals \cite{Roy}, the effect of anion Frenkel defects on the endurance behaviour of Cu doped SnO$ _{2} $ based memristors \cite{Meifang}, \textit{etc}. On doping Co in SnO$ _{2} $ nanocrystals \cite{Roy}, it was observed that the concentration of intrinsic Sn-related non-stoichiometric point defect Sn$ _{i} $ increases whereas that of O-related Frenkel stoichiometric point defects -- \textit{V}$ _{\text{O}} $ and O$ _{i} $ decreases. The subsequent effect on the structural and optical properties was such that a local symmetry breaking in the nanocrystals exist; spectral contribution of UV-emission increased till 0.5 at.\% Co doping and then it decreased while the green emission followed an opposite trend, with no conclusive trend for blue and violet emissions. The ionic radius mismatch and charge imbalance occuring in the host lattice on Co doping was considered to be the origin for such interesting properties. Since, both Ni and Co act as stable dopants in SnO$ _{2} $ and have been found to exist in mixed valence states \cite{Roy, Srinivas}, hence, Ni when co-doped with Co in SnO$ _{2} $ nanocrystals can be expected to manipulate the intrinsic point defects of the host lattice such that the structural and optical properties would be influenced in a way much different than that for only Co doped SnO$ _{2} $ nanocrystals. Some reports on Fe-Ni \cite{Kiyoshi_Fe-Ni} and Fe-Co \cite{Nomura_Fe-Co} co-doped SnO$ _{2} $ nanocrystals are available, but reports on Co-Ni co-doped SnO$ _{2} $ nanocrystals are still absent. Also, although local symmetry breaking in the SnO$ _{2} $-based nanocrystals has been found to cause many interesting properties, like, otherwise forbidden UV-emission \cite{Roy}, the mechanism of local symmetry breaking is unclear, but it can be expected to be associated with the effect of point defects on Sn -- O octahedral coordination geometry. As such, an understanding of the formation of intrinsic defects and their effect on the geometry of Sn -- O octahedron and other physical properties becomes very important, a detailed study of which is lacking for any co-doped system. 

Here, we report on the identification of intrinsic point defects on Co-Ni co-doping at Sn site in SnO$ _{2} $ nanocrystals and their effect on structural, specially the octahedral coordination geometry of Sn and the trigonal planar coordination geometry of O, and optical properties of the nanocrystals. To the best of our knowledge, this is the first experimental study for the identification and corresponding effect of intrinsic point defects on co-doping cobalt and nickel at Sn site in the SnO$ _{2} $ nanocrystals. The results will help to understand the origin and nature of the optical, electronic and magnetic properties of the co-doped nanocrystals.       

%-------------------------------------------------------------------------------------------------------------------------------------------------------------------------------------
\section{Experimental Details}

\subsection{Synthesis} 
Sn$ _{0.97-y} $Co$ _{0.03} $Ni$  _{y}$O$ _{2} $ (with; 0.00 $ \leq $ \textit{y} $ \leq $ 0.04) samples have been synthesised in an oxygen rich environment by salt hydrolysis of chloride precursors using chemical co-precipitation method. The samples have been designated as SnO$ _{2} $, and Ni0, Ni0.5, Ni2, Ni3 and Ni4 for nickel co-doping concentration as \textit{y} = 0.00, 0.005, 0.02, 0.03, 0.04 respectively. The synthesis procedure has been followed as reported earlier \cite{Roy}. NiCl$ _{2}$.6H$ _{2} $O (HIMEDIA, min. assay = 97\% ) was used as a precursor for Ni in addition to the precursors for Sn (SnCl$ _{4} $.5H$ _{2} $O; MOLYCHEM, min. assay = 98\%) and Co (CoCl$ _{2}$.6H$ _{2} $O; HIMEDIA, min. assay = 99\%), in stoichiometric amounts as per the co-doping concentration \textit{y}, to prepare the 0.123 M homogenous aqueous precursor solution. All other synthesis conditions have been kept same as earlier \cite{Roy}. 

\subsection{Characterisation studies \& measurements}
\textit{Structural characterisations}: The structural characterisations of the samples have been done using X-ray  diffraction (XRD), transmission electron microscopy (TEM) and X-ray photoelectron spectroscopy (XPS). Room temperature X-ray diffractograms of the samples were recorded using a Miniflex - II (Rigaku, Japan) X-ray diffractometer with Cu - K$ _{\alpha} $ radiation ($ \lambda $ = 1.5406 \AA) in Bragg-Brentano geometry. The loose powder samples were levelled on a sample holder for ensuring a smooth surface before recording the diffractograms. Low resolution TEM, high resolution-TEM (HR-TEM) and selected area electron diffraction (SAED) measurements of the samples were carried out using a FEI Tecnai G$^{2}$ 20 TWIN model (Japan) electron microscope. The samples were well dispersed in ethanol by ultrasonication for 15 minutes and then coated on a 300 mesh carbon-coated copper grid for the measurements. XPS (survey, Sn3d core level and O1s core level) scans of the Ni0, Ni2 and Ni4 samples have been performed at room temperature using monochromatic Al-K$ _{\alpha} $ line at 1486.7 eV (300 Watt, 15 kV) from Omicron multiprobe surface analysis system (Scienta Omicron GmbH), operating at an average vacuum of 2.5 $  \times$ 10$ ^{-9} $ Torr.

\textit{Study of optical properties}: The effect of co-doping on the optical properties of the samples were studied using reflectance spectroscopy and photoluminescence (PL) spectroscopy. The reflectance spectra of the samples were obtained at room temperature using a V750 (Jasco International Inc., Japan) dual beam UV-Vis spectrophotometer. The spectra were collected in the wavelength range of 200 -- 800 nm at 0.5 nm data intervals with a scan speed of 40 nm/ min and 0.5 nm bandwidth. The response time of the photomultiplier tube (the detector) was kept at 0.96 sec for the whole range of measurements. The PL emission spectra were collected at room temperature using a LS-45 (PerkinElmer, U.S.A.) fluorescence spectrometer. The well dispersed samples in distilled water were excited at an excitation wavelength of 310 nm and then scanned from 320 nm to 600 nm at a scan speed of 100 nm/min and a step scan of 0.5 nm for recording the emission spectra.

\textit{Study of magnetic property}: Room temperature response of magnetisation of the nanocrystals towards the applied magnetic field was studied using an EZ9 vibrating sample magnetometer (VSM; Microsense, U.S.A.). The powder samples of the nanocrystals were filled in acrylic cups, which were then mounted on a quartz sample holder for the measurements. Prior to each measurement, magnetisation vs. applied magnetic field data were recorded for blank acrylic cups, which were subtracted from the corresponding data of the samples for the necessary background corrections. The temperature dependent magnetisation measurements for the Ni0 and Ni3 nanocrystals were done using MPMS - 03 (Quantum Design International, U.S.A.) SQUID  magnetometer.

%-----------------------------------------------------------------------------------------------------------------------------------------------------------------------------------------

\section{Results}
\subsection{X-ray diffraction (XRD)}
%---------------------------------------
\begin{figure}[ht]
\centering
\includegraphics[width=1.00\linewidth]{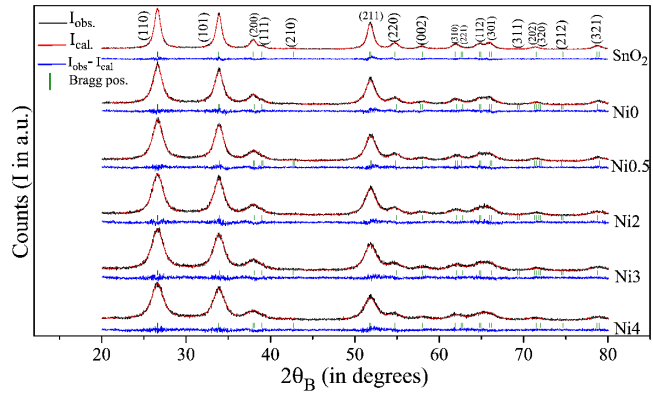}
\caption{XRD profiles, along-with Rietveld refinement, of the nanocrystals. The black lines correspond to the observed intensity (I$ _{\text{obs.}} $), the red lines correspond to the calculated intensity (I$ _{\text{cal.}} $), horizontal blue lines represent the difference between observed and calculated intensities (I$ _{\text{obs.}} $--I$ _{\text{cal.}} $) and the green vertical ticks correspond to the Bragg positions.}
\label{figxrd}
\end{figure}
%---------------------------------------
\begin{figure}[ht]
\centering
\includegraphics[width=1.00\linewidth]{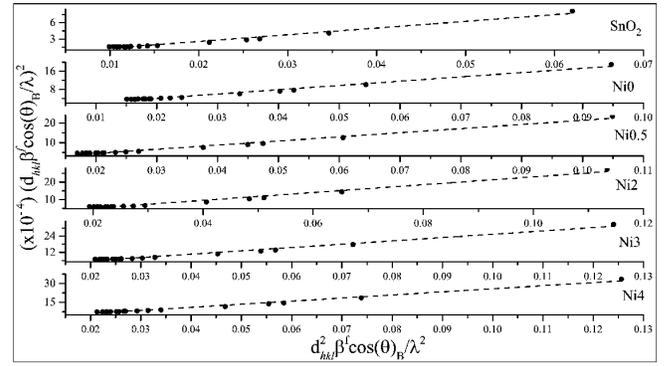}
\caption{Size-strain plot for the nanocrystals. The dotted lines represent the linear fits corresponding to the size-strain plots\cite{Roy}.}
\label{figssp}
\end{figure}
%---------------------------------------
\begin{figure}[ht]
\centering
\includegraphics[width=1.00\linewidth]{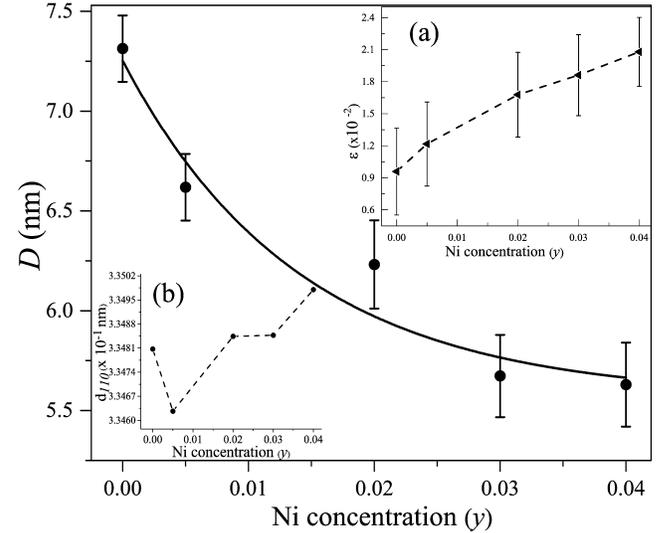}
\caption{The variation of average crystallite size (\textit{D}) with Ni co-doping concentration (\textit{y}) for the nanocrystals. The inset (a) shows the variation of average lattice strain ($ \varepsilon $) and inset (b) shows the variation of d$_{\textit{110}}$with Ni co-doping concentration (\textit{y}). The solid line in the figure denotes the corresponding exponential fit and the dashed line is eye-guide.}
\label{figsize}
\end{figure}
%-----------------------------------------
\begin{figure}[ht]
\centering
\includegraphics[width=1.00\linewidth]{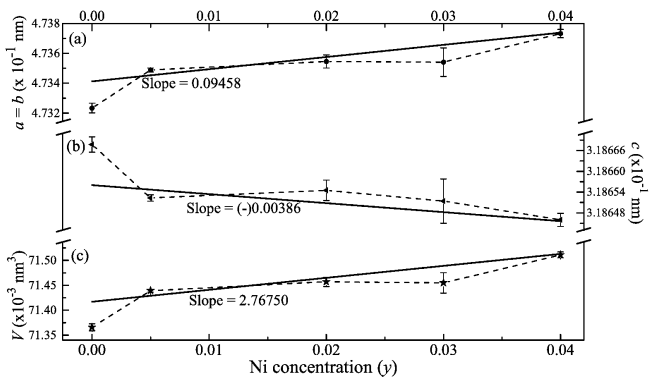}
\caption{The variation of lattice parameter \textit{a} = \textit{b} (a); \textit{c} (b); and volume \textit{V}(c); of the nanocrystals with Ni co-doping concentration (\textit{y}). The solid lines, representing the corresponding linear fit, and the dashed lines serve as eye guides.}
\label{figlattice}
\end{figure}
%------------------------------------------
The XRD patterns of the nanocrystals, along-with corresponding Rietveld refinements, are shown in Fig. \ref{figxrd}. The XRD pattern of nanocrystalline SnO$ _{2} $ (with an average crystallite size of 11.25 nm) is also shown in Fig. \ref{figxrd} for the sake of comparison. Rietveld refinement of the XRD patterns was done using EdPCR program (version 2.00) of FullProf Suite (version July -- 2017). The refinement data so obtained was used to calculate the structural parameters of the samples using BondStr program (version July -- 2010) of FullProf Suite and VESTA (Visualization for Electronic and Structural Analysis, Ver. 3.4.4). All the peak positions are indexed in accordance with JCPDS file\# 41-1445 for tetragonal SnO$ _{2} $. From the refinement patterns, it can be observed that the Sn$ _{0.97-y} $Co$ _{0.03} $Ni$ _{y} $O$ _{2} $ nanocrystals are of single phase and correspond to the P4$ _{2} $/\textit{mnm} space group of tetragonal cassiterite SnO$_{2}$. 

The average crystallite size (\textit{D}) and the average lattice strain ($ \varepsilon$) developed in the nanocrystals on co-doping have been calculated using size-strain plots (SSP) \cite{Prince} (Fig. \ref{figssp}), following a method reported earlier \cite{Roy}. Figure \ref{figsize} shows the variation of \textit{D} with \textit{y} and the inset (a) of Fig. \ref{figsize} shows the corresponding variation in $ \varepsilon$.  It can be observed that \textit{D} exponentially decreases from $ \approx $ 7.31 nm for Ni0 nanocrystals to $ \approx $ 5.63 nm for Ni4 nanocrystals, with increase in Ni concentration (\textit{y}), however, the corresponding variation of $ \varepsilon$ is quite linear as well as is tensile in nature. Figure \ref{figlattice} shows the variation of lattice parameters \textit{a} (= \textit{b}) and \textit{c} with Ni concentration (\textit{y}). With increase in \textit{y}, \textit{a} (= \textit{b}) slightly increases (Fig. \ref{figlattice}(a)) and \textit{c} slightly decreases (Fig. \ref{figlattice}(b)), with an increase in total volume \textit{V} (= \textit{a}$^{2}$\textit{c}) of the unit cell (Fig. \ref{figlattice}(c)). Although the variation in the lattice parameters is very small, but a different response of \textit{a} (= \textit{b}) as compared to \textit{c} towards \textit{y} can be readily observed from Fig. \ref{figlattice}.

%-----------------------------------------------------------------------------------------------------------------------
\subsection{Transmission electron microscopy (TEM)}
%---------------------------------------------------------------
\begin{figure}[ht]
\centering
\includegraphics[width=1.00\linewidth]{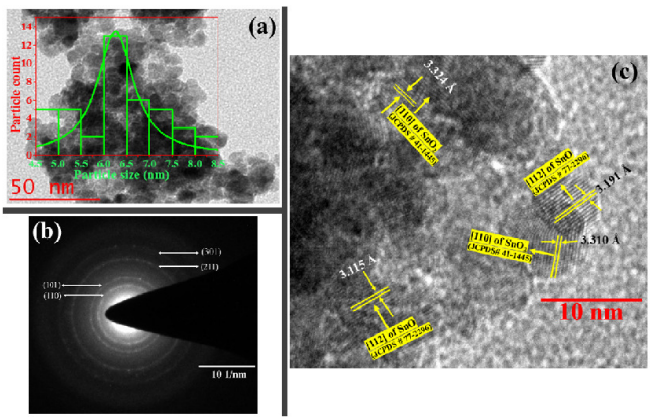}
\caption{Bright field low-resolution TEM (a); selected area electron diffraction (SAED) pattern (b); and high-resolution TEM (HR-TEM) image (c) of the Ni3 nanocrystals.}
\label{figtem}
\end{figure}
%---------------------------------------------------------------
Transmission electron microscopy (TEM) provides a precise estimate of the morphology and microstructure of the nanocrystals. Figure \ref{figtem}(a) shows the bright field low-resolution TEM images, along-with particle size distribution for Ni3 nanocrystals. 0.5 nm bin size has been used for investigating the distribution. Most of the particles are spherical in shape and are in the size range of 6 nm to 6.5 nm, with the most probable particle size (\textit{D$ _{\text{TEM}} $}) being around 6.25 nm. This corroborates the average crystallite size (\textit{D}) (Fig. \ref{figsize}) as estimated from X-ray diffractogram of the nanocrystals. 

Selected area electron diffraction (SAED) patterns is considered to give a better insight into the crystalline nature of nanocrystals. Circular ring-like patterns of very closely spaced bright spots are observed from the SAED pattern (Fig. \ref{figtem}(b)) of the present Ni3 nanocrystals, which indicates good crystallinity of the nanocrystals. Indexing of the SAED patterns (Fig. \ref{figtem}(b)) corresponds to the same order of lattice planes as obtained from Rietveld refinement of X-ray diffractogram of the nanocrystals, i.e., they match well with that of tetragonal SnO$  _{2}$-phase.  

For having a view of the atomic arrangement in the nanocrystals, high-resolution TEM (HR-TEM) of the nanocrystals have been performed, as shown in Fig. \ref{figtem}(c). Intertwined atomic planes can be observed, which indicates the presence of more than one type of atomic arrangement in the nanocrystals. From the Fig. \ref{figtem}(c), four typical sets of atomic planes have been identified -- with interplanar spacings (d$ _{hkl} $) of 3.310 \AA (Region I), 3.324 \AA (Region II), 3.191 \AA (Region III) and 3.115 \AA (Region IV). The d$ _{hkl} $ of $ \approx $ 3.3 \AA (corresponding to Region I \& II) matches with that of d$ _{110} $ of the Ni3 nanocrystals as obtained from Rietveld refinement (d$ _{110} $ = 3.34807 \AA) considering the tetragonal SnO$ _{2} $ phase (JCPDS\# 41-1445), where Sn is in +4 oxidation state. However, the d$ _{hkl} $ of $ \approx $ 3.1 \AA (corresponding to Region III \& IV) do not match with any (\textit{hkl}) value of tetragonal SnO$ _{2} $, but, they match with that of \textit{$\lbrace112\rbrace$} planes of orthorhombic SnO phase (JCPDS\# 77-2296), where Sn is in +2 oxidation state. Thus, contrary to diffraction results (XRD and SAED), HR-TEM indicates formation of some local structures resembling SnO in the overall tetragonal SnO$ _{2} $ lattice of the nanocrystals. This contradiction to diffraction results can be expected since, the average particle size of the nanocrystals are very small ($ \approx $ 6.25 nm) as well as with not-so-good crystallinity of the nanocrystals, the diffraction patterns cannot provide accurate structural information for the nanocrystals \cite{Heather, Dong}.        
 
%------------------------------------------------------------------------------------------------------------------------
\subsection{X-ray photoelectron spectroscopy (XPS)}
%---------------------------------------------------------
\begin{figure}[ht]
\centering
\includegraphics[width=1.00\linewidth]{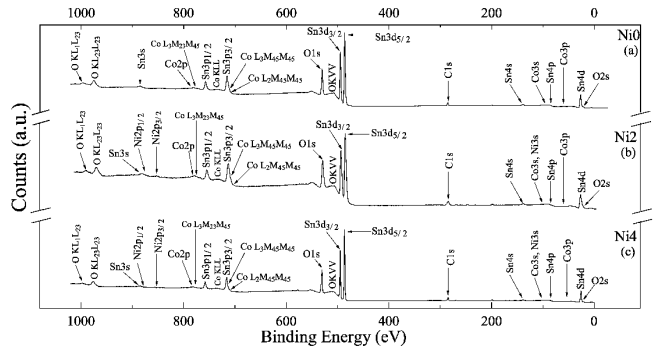}
\caption{Room temperature survey scan XPS spectra of the nanocrystals.}
\label{figxpssurvey}
\end{figure}
%--------------------------------------------------------
\begin{figure}[hb]
\centering
\includegraphics[width=1.00\linewidth]{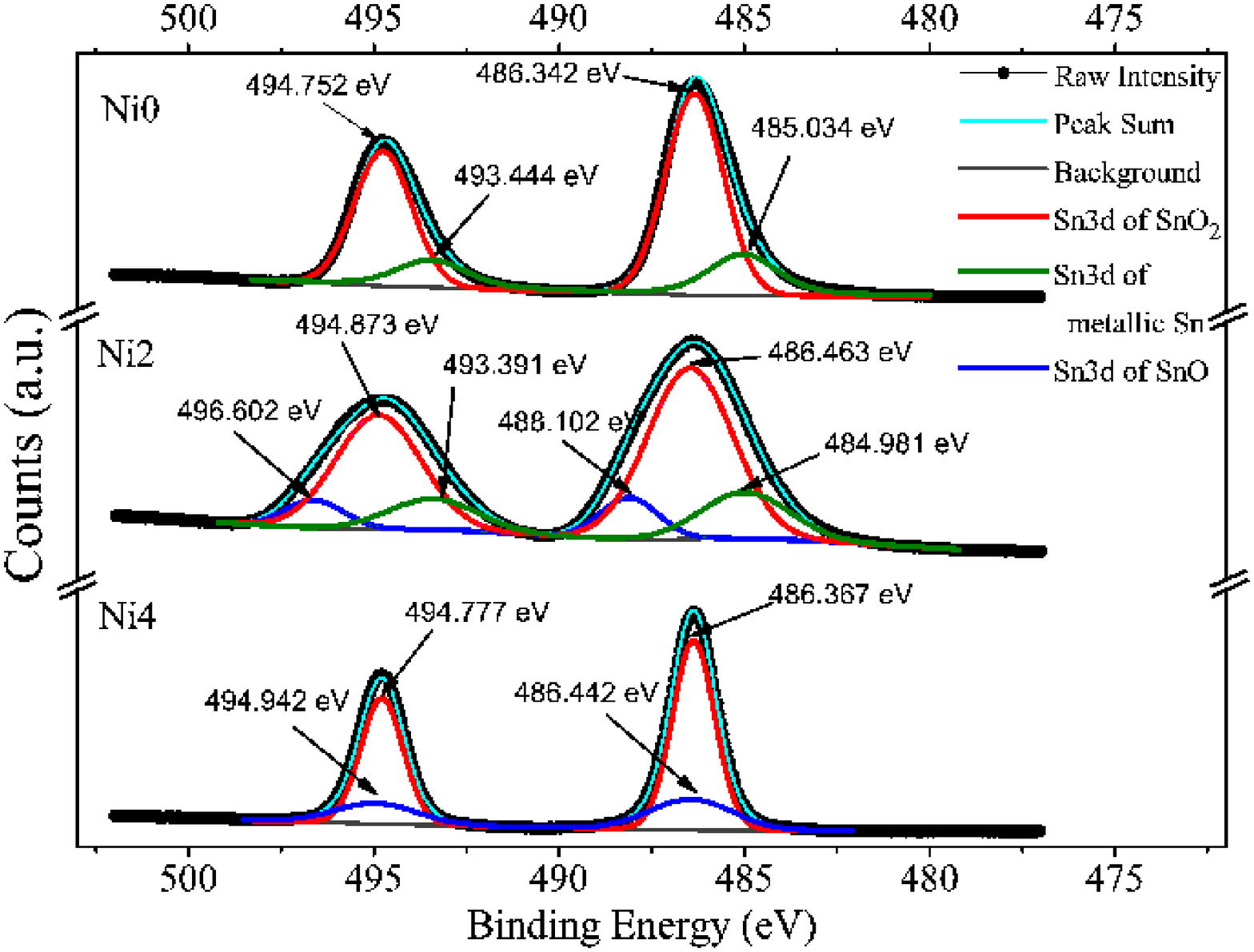}
\caption{Room temperature Sn3d core level XPS spectra of the nanocrystals.}
\label{figsn3d}
\end{figure}
%--------------------------------------------------------
\begin{figure}[ht]
\centering
\includegraphics[width=1.00\linewidth]{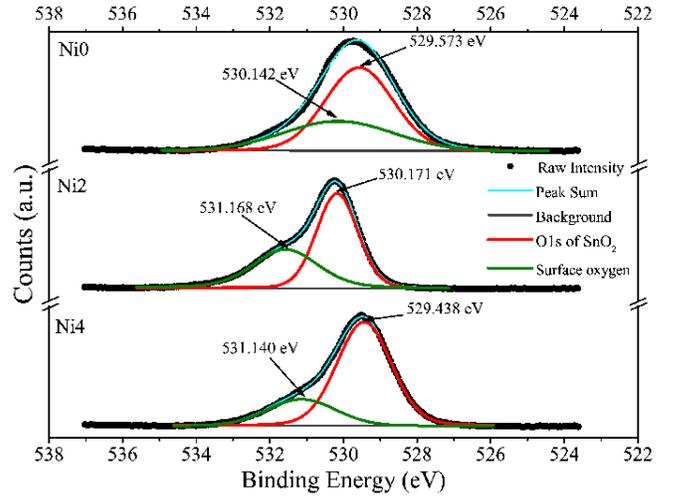}
\caption{Room temperature O1s core level XPS spectra of the nanocrystals.}
\label{figo1s}
\end{figure}
%-------------------------------------------------------
Room temperature XPS study of the nanocrystals can be useful to determine the elemental composition as well as electronic states of Sn and O in the samples. All the observed peak positions, along-with the obtained spin-orbit splittings (SOS), in the XPS survey and core-level scans of the present nanocrystals have been indexed from the National Institute of Standards and Technology (NSIT) XPS database \cite{NIST}. The experimental peak area values were first divided by the number of sweeps to get the atomic areas, which was then divided by the respective atomic scattering factors (ASF) for the quantitative analysis of the samples. C1s signal at 284.5 eV has been considered as standard for calibrating the spectral positions to encompass the sample charging effects.

The XPS survey scans for the Ni0, Ni2 and Ni4 nanocrystals are shown in Fig. \ref{figxpssurvey}. Observed peaks confirm the presence of Sn, O, Co and Ni in the nanocrystals and that the nanocrystals do not have an admixture of any other element.

The core level  XPS spectra of Sn3d (Fig. \ref{figsn3d}) show the presence of Sn3d doublets 3d$ _{3/2} $ and 3d$ _{5/2} $ centred at around 494.7 eV and 486.3 eV respectively. Clear asymmetry in the peaks can be observed, indicating presence of more than one peak in the region. To clarify it, the Sn3d doublets have been deconvoluted (Fig. \ref{figsn3d}). It has been observed that the deconvolution of Sn3d$ _{5/2} $ peak lead to two peaks for Ni0 and Ni4 nanocrystals and three peaks for Ni2 nanocrystals. The best fit results have been obtained with a spin-orbit splitting (SOS) value of 8.41 eV for both the deconvoluted peaks of 3d$ _{5/2} $ in case of Ni0, whereas for Ni4, the corresponding best fit have been obtained for SOS values of 8.41 eV and 8.5 eV respective to each peak. For Ni2, two peaks are fitted with SOS of 8.41 eV and the third peak is fitted with an SOS of 8.5 eV. For Ni0, it is observed from Fig. \ref{figsn3d} that one of the deconvoluted peaks corresponds to the B.E. of 486.342 eV in the 3d$ _{5/2} $ region and 494.752 eV in the 3d$ _{3/2} $ region. From the observed B.E. and SOS of 8.41 eV, the peak can be assigned to lattice Sn of SnO$ _{2} $ \cite{NIST}. Deconvolution of the 3d peaks for Ni2 and Ni4 samples also revealed a presence of similar peaks. Upon careful examination of the 3d$ _{5/2} $ peak at 485.034 eV for Ni0 and at 484.981 eV for Ni2 nanocrystals (Fig. \ref{figsn3d}), both with an SOS value of 8.41 eV, it can be found that it corresponds to that of Sn in metallic state \cite{NIST}. Furthermore, the 3d$ _{5/2} $ peak at 488.102 eV for Ni2 and 486.442 eV for Ni4 were fitted with an SOS value of 8.5 eV (Fig. \ref{figsn3d}), which is only assigned to Sn3d of SnO \cite{NIST}. Hence, this peak can be assigned to the +2 oxidation state of Sn in SnO. In all the nanocrystals, no such significant chemical shift for any of the Sn3d peaks (Fig. \ref{figsn3d}) has been observed.  

The O1s core level XPS spectra (Fig. \ref{figo1s}) show the presence of O1s singlet peak centred at $ \approx $ 529.8 eV for the nanocrystals. A clear asymmetric broadening of the peak on the higher B.E. side has been observed for all the nanocrystals. Deconvolution of the peak yields two symmetric peaks centred at 529.573 eV and 530.142 eV for Ni0; at 530.171 eV and 531.168 eV for Ni2; and at 529.438 eV and 531.140 eV for Ni4. The peak at $ \approx $ 529.573 eV can be assigned to lattice O in SnO$ _{2} $ \cite{NIST}. Also, there is no significant change in the B.E. value of this peak with Ni co-doping concentration (\textit{y}). The peak at $ \approx $ 530.142 eV is assigned to the surface dangling bond states (E$ _{s} $) resulting from the species, such as, O$ _{2}^{2-} $ and O$ ^{-} $\cite{Roy}. 

%-----------------------------------------------------------------------------------------------------------------------
\subsection{Reflectance spectroscopy}
%-----------------------------------------------
\begin{figure}[ht]
\centering
\includegraphics[width=1.00\linewidth]{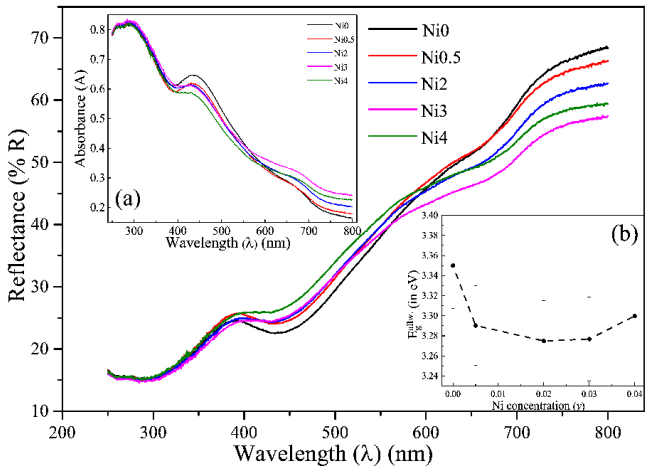}
\caption{Room temperature reflectance spectra of the nanocrystals. The inset (a) of the figure represents the corresponding absorbance spectra of the nanocrystals; whereas the inset (b) of the figure shows the variation of allowed optical band gap (E$ _{\text{g}}^{\text{allw.}} $) of the nanocrystals, the dashed line being served as eye guide.}
\label{figreflectance}
\end{figure}
%---------------------------------------------------------
The reflectance spectroscopy is an important tool to study the optical absorption process in semiconductor nanocrystals which convey many important informations regarding the nature of optical transitions (direct or indirect), the optical band gap, lattice imperfections, etc. The reflectance spectra of the Sn$ _{0.97-\text{y}} $Co$ _{0.03} $Ni$ _{\text{y}} $O$ _{2} $ nanocrystals as a variation of incident photon wavelength ($ \lambda $) is shown in Fig. \ref{figreflectance}. SnO$ _{2} $ in its bulk state is a direct band gap semiconductor \cite{Frohlich}. However, due to the presence of an inversion centre symmetry in the ideal SnO$ _{2} $ crystals, the band to band edge transitions are of forbidden nature owing to the even parity symmetry of the band edge quantum states \cite{Agekyan}. It has been reported that reducing the size of SnO$ _{2} $ crystals to nano regime, inducing defects by doping, \textit{etc.} can modify the symmetry of the band edge quantum states such that the forbidden nature of direct optical transitions can be suppressed partially or completely \cite{Y_Feng}. The direct allowed band gap (E$ _{\text{g}}^{\text{allw.}} $) of the nanocrystals have been calculated from the reflectance spectra using Tauc's plot following the method described elsewhere \cite{Roy}. The existence of direct allowed transitions in the nanocrystals establishes a modification of the symmetry of the band edge quantum states. This symmetry modification has also been reported earlier for Co doped SnO$  _{2}$ nanocrystals \cite{Roy}. The variation of E$ _{\text{g}}^{\text{allw.}} $ with \textit{y} is shown in inset (b) of Fig.\ref{figreflectance}. It can be observed that the optical band gap of the nanocrystals first decreases with increase in \textit{y} upto \textit{y} = 0.02 and then increases thereafter. This is different from the earlier observation for only Co doped SnO$  _{2}$ nanocrystals \cite{Roy}, where a red shift of the band gap has been observed throughout the Co doping concentration, confirming the effect of Ni co-doping with Co on the optical band gap modification of the SnO$  _{2}$ nanocrystals.  
%----------------------------------------------
\begin{figure}[ht]
\centering
\includegraphics[width=1.00\linewidth]{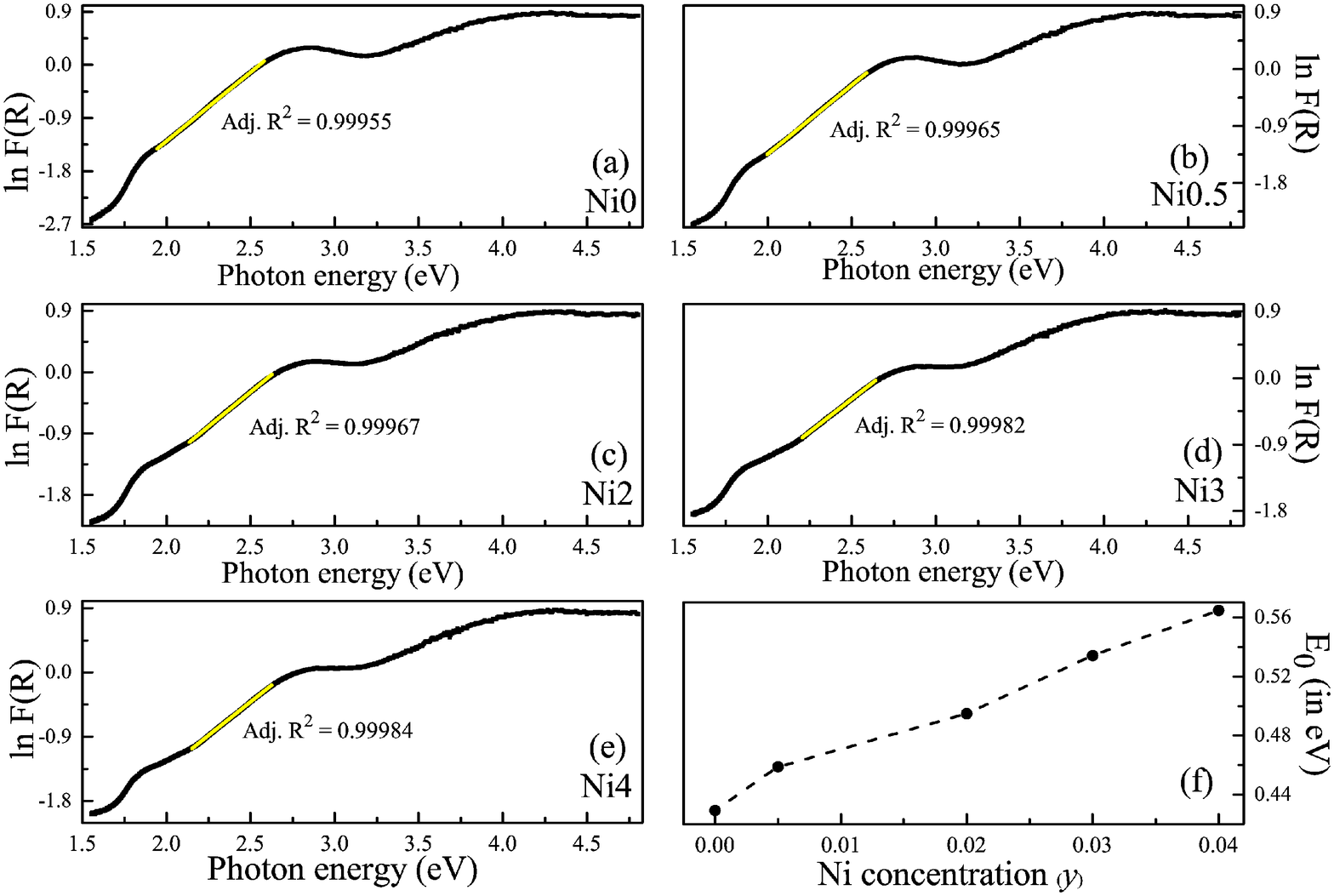}
\caption{The plots for Urbach type absorption (according to eqn. (\ref{eqnurbach}) in the nanocrystals ((a)-(e); variation of the E$_0$ with Ni co-doping concentration (\textit{y}) (f). The solid yellow lines in (a) -- (e) represent the corresponding linear fits whereas the dotted line in (f) is for eye guide. }
\label{figurbach}
\end{figure}
%----------------------------------------------

Additionally, an exponential edge, i.e., the Urbach edge \cite{urbach} has also been observed near the optical band edge in the reflectance spectra of the nanocrystals. The Urbach type absorption is represented by \cite{Pankove, Mott, Tauc, ghoshprb, Sajeev}:
\begin{equation} \label{eqnurbach}
\alpha(\text{h}\nu) = \alpha_{0}\text{exp}\left[ \gamma\left\lbrace \dfrac{\text{h}\nu - \text{h}\nu _{0}}{\text{k}_{\text{B}}T}\right\rbrace \right] 
\end{equation}
where, $ \alpha (\text{h}\nu)$ is the absorption coefficient of the nanocrystals at an incident photon energy h$ \nu $; $ \alpha_{0} $, $ \gamma $ and h$ \nu_{0} $ are fitting parameters and material-dependent. The empirical parameter $ \frac{\text{k}_{\text{B}}T}{\gamma} $ has the dimensions of energy. From eqn. (\ref{eqnurbach}), it can be shown that \cite{Pankove}:
\begin{equation} \label{eqnpankove1}
\text{E}_{0} = \left[ \dfrac{\text{d} (\text{ln}(\alpha)) }{\text{d} (\text{h}\nu) }\right] ^{-1}
\end{equation}
where, E$ _{0} $ = $ \frac{\text{k}_{\text{B}}\text{T}}{\gamma} $. Substituting Kubelka-Munk function (F(R)) for $ \alpha $ in eqn. (\ref{eqnpankove1}) (since, F(R) is proportional to $ \alpha(\text{h}\nu) $, we have:
\begin{equation} \label{eqnpankove2}
\text{E}_{0} = \left[ \dfrac{\text{d} (\text{ln}(\text{F(R)})) }{\text{d} (\text{h}\nu) }\right] ^{-1}
\end{equation}
The plot of lnF(R) vs. h$ \nu $ for the present nanocrystals (Fig. \ref{figurbach}(a)-(e)) yield a straight line for photon energies below the optical band gap of the nanocrystals, confirming the Urbach type absorption \cite{urbach}. This Urbach type absorption occurs when the electronic density of states exponentially tail into the band gap region owing to various structural disorders in the lattice of the semiconductors \cite{Sajeev}. The empirical parameter E$ _{0} $ describes the distribution of density of states in band structure of the nanocrystals and is correlated with impurity concentration responsible for disorder in crystallinity of the nanocrystals, leading to the Urbach edge \cite{Pankove, ghoshprb}. Urbach type absorption has also been found earlier for SnO$ _{2} $-based nanostructures \cite{Nagasawa, Soumen, J_Jiang} as well as for other semiconductors \cite{ghoshprb}. In fact, a strong Urbach type absorption has been observed for Co doped SnO$ _{2} $ nanocrystals reported earlier \cite{Roy}, and was found to be pronounced on Co doping. E$ _{0} $ for the present nanocrystals has been calculated, as according to eqn. (\ref{eqnpankove2}) from the inverse of the slope of the straight lines in Fig. \ref{figurbach}(a)-(e) and its variation with Ni co-doping concentration (\textit{y}) is shown in Fig. \ref{figurbach}(f). On co-doping Ni with Co, i.e., with increase in \textit{y}, E$_{0}$ increases indicating an enhancement of structural disorders in lattice of the SnO$ _{2} $ nanocrystals because of co-doping. 

%------------------------------------------------------------------------------------------------------------------------
\subsection{Photoluminescence (PL) spectroscopy}
%------------------------------------------------------
\begin{figure}[ht]
\centering
\includegraphics[width=1.00\linewidth]{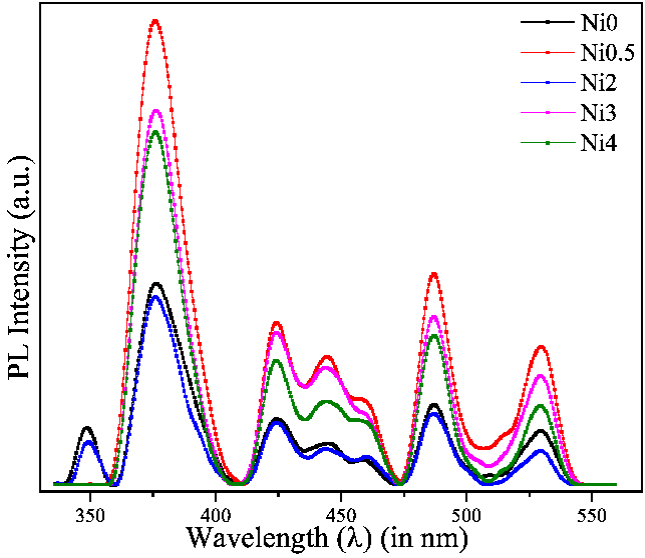}
\caption{Room temperature PL spectra of the nanocrystals.}
\label{figpl}
\end{figure}
%------------------------------------------------------
\begin{figure}[hb]
\centering
\includegraphics[width=1.00\linewidth]{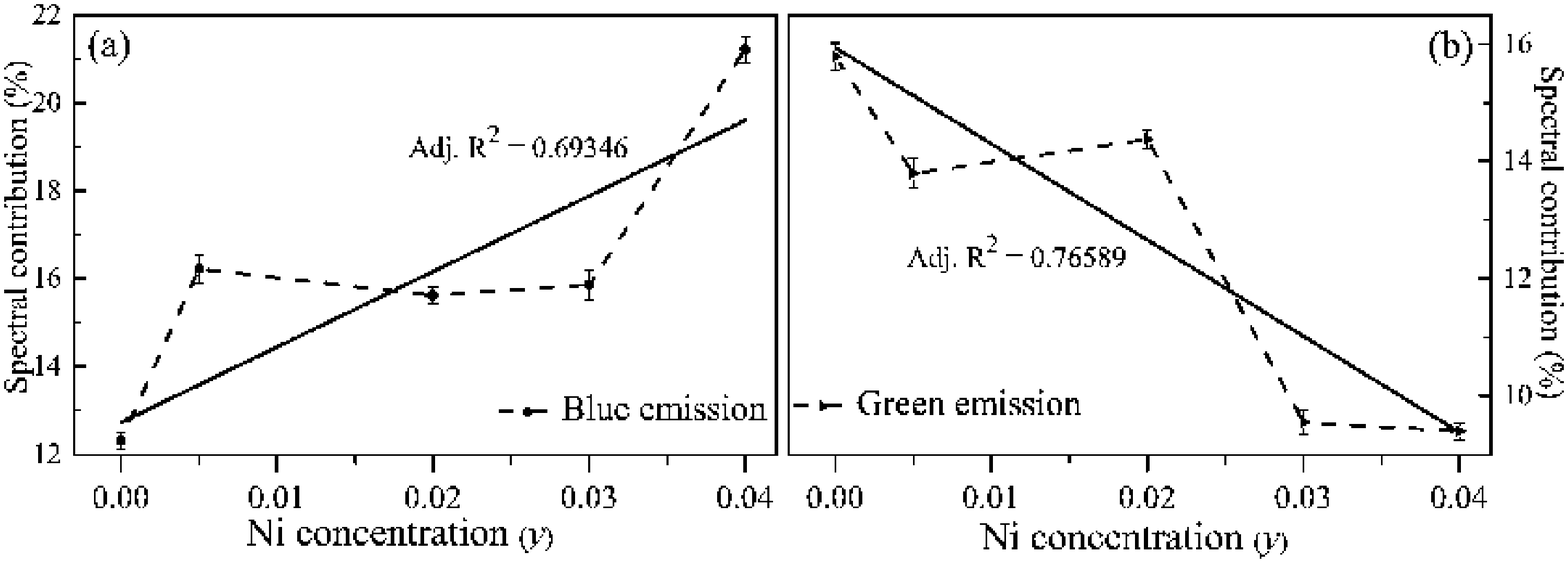}
\caption{The variation of spectral contribution (\%) of blue emission (a); and green emission (b) in the total emission spectra of the nanocrystals, with Ni co-doping concentration (\textit{y}). The solid lines, representing the linear fits and the dotted lines are eye guides.}
\label{figspectral}
\end{figure}
%-------------------------------------------------------
Photoluminescence (PL) spectroscopy is a non-destructive spectroscopic technique useful for studying the presence and corresponding effect of various intrinsic and extrinsic point defects in semiconductor nanocrystals. The fluorescence PL spectra of the present nanocrystals is shown in Fig. \ref{figpl}. The observed PL spectra is divided into four spectral regions -- ultraviolet (UV) (below 380 nm), violet (380 nm to 450 nm), blue (450 nm to 490 nm) and green (490 to 540 nm) emissions. The presence of these emission peaks at energies corresponding to sub-band gap region indicate the presence of various point defect related energy levels in the forbidden zone, acting as recombination levels for electrons relaxing from conduction band to valence band.   

To obtain the spectral contribution of blue and green emissions towards the total emission intensity, all the observed PL emission peaks in Fig. \ref{figpl} has been deconvoluted and fitted using Gaussian function (eqn.(\ref{eqngaussian})):
\begin{equation} \label{eqngaussian}
I=\frac{A}{\sqrt{2\pi}\beta}exp\left[ \frac{-1}{2}\left( \frac{\lambda-\lambda_{0}}{\beta}\right) ^{2}\right] 
\end{equation}
where; \textit{I} is the observed PL intensity; $ \lambda $ is the emitted photon wavelength; \textit{A} is the Gaussian area of the peaks, called as peak intensity; $ \lambda_{0} $ is the obtained peak positions on the wavelength scale; and $ \beta $ is the peak width. The so obtained peak intensities for a nanocrystal were added to obtain the total PL emission intensity. The spectral contribution percentage for blue and green emissions in the total emission spectrum for a nanocrystal was then calculated by taking the ratio of respective peak intensities with the total PL emission intensity. Figure \ref{figspectral} shows the variation of spectral contribution of the blue and green emissions to the total emission spectra for a nanocrystal w.r.t. the Ni co-doping concentration (\textit{y}). It can be observed that the spectral contribution of blue luminescence (Fig. \ref{figspectral}(a)) shows a remarkable enhancement of $ \approx $ (72.33 $ \pm $ 1.74)\% for Ni4 w.r.t. Ni0 nanocrystals, whereas the corresponding spectral contribution for green emission (Fig. \ref{figspectral}(b)) has decreased by $ \approx $ (40.49 $ \pm $ 1.46)\%. 

%-----------------------------------------------------------------------------------------------------------------------
\subsection{Magnetic property}
%-----------------------------------------------
\begin{figure}[ht]
\centering
\includegraphics[width=1.00\linewidth]{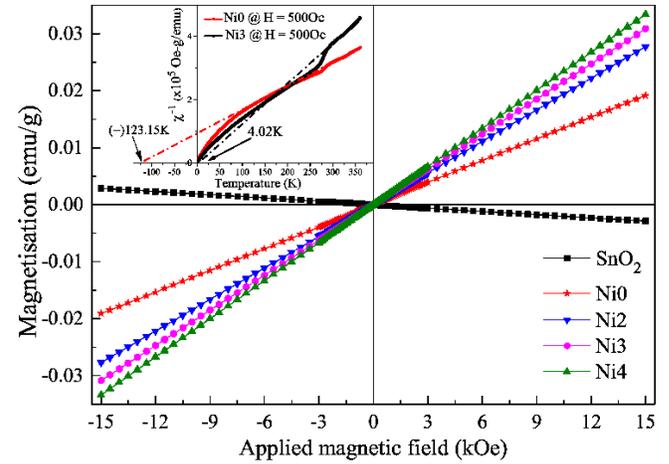}
\caption{Room temperature response of the magnetisation (M) towards the applied magnetic field (H) for the nanocrystals. The inset figure shows the temperature dependent magnetisation (M) curves for the Ni0 and Ni3 nanocrystals. The dashed-dot lines in the inset figure represents the corresponding Curie-Weiss fits.}
\label{figmagnetic}
\end{figure}
%----------------------------------------------
The variation of magnetisation (M) with applied magnetic field (H) at 300 K for the present nanocrystals is shown in Fig. \ref{figmagnetic}. Pure SnO$ _{2} $ nanocrystals exhibit diamagnetism, as evident from the black curve in Fig. \ref{figmagnetic}. However, on co-doping with Ni and Co, the M-H behaviour of the nanocrystals seem to correspond to paramagnetism at 300 K. To confirm the dominant magnetic interactions, magnetisation (M) as a function of temperature (T) of the nanocrystals is studied at an applied magnetic field of 500 Oe. The resulting variation of inverse susceptibility ($ \chi^{-1} $) with temperature (T) is shown in the inset of Fig. \ref{figmagnetic}. Now, a system, above a magnetic ordering temperature, is expected to follow the Curie-Weiss law (eqn. (\ref{eqncurieweiss})):
\begin{equation} \label{eqncurieweiss}
\chi = \dfrac{\text{C}}{\text{T}-\Theta}
\end{equation}
where, C is the Curie constant; and $ \Theta $ is the Curie-Weiss temperature. Thus, a plot of $ \chi^{-1} $ vs. T, i.e., eqn. (\ref{eqncurieweissplot}) 
\begin{equation} \label{eqncurieweissplot}
\chi^{-1} = \dfrac{\text{T}}{\text{C}}- \dfrac{\Theta}{\text{C}}
\end{equation}
is expected to give a linear region in the high temperature range with a zero, negative and positive intercept on the $ \chi^{-1} $-axis respectively for dominant paramagnetic, ferromagnetic and antiferromagnetic interactions in the system. From the inset of Fig. \ref{figmagnetic}, it can be seen that a linear region exists in the high temperature range of 290 K -- 360 K for both the Ni0 and Ni3 nanocrystals, confirming to Curie-Weiss law. The linear fit using eqn. (\ref{eqncurieweissplot}) for the nanocrystals yields a positive  intercept with $ \Theta \approx$ (-)123.15 K for Ni0 nanocrystals, whereas that for Ni3 nanocrystals a negative intercept is obtained with $ \Theta \approx$ 4.02 K. The observed $ \Theta \approx$ (-)123.15 K for Ni0 nanocrystals is sufficiently high enough to consider the presence of dominant antiferromagnetic interactions in the nanocrystals. However, the absence of any noticeable ferromagnetic loop (in Fig. \ref{figmagnetic}) as well as the sufficiently low $ \Theta \approx$ 4.02 K in Ni3 nanocrystals suggests that the samples are very weakly ferromagnetic. This conversion from dominant antiferromagnetic interactions to feeble ferromagnetic interactions suggests the inclusion of paramagnetic centres in the lattice of nanocrystals because of co-doping. However, the exact origin of magnetic interactions in the nanocrystals need to be investigated further.

%----------------------------------------------------------------------------------------------------------------------------------------------------------------------------------------------------------
\section{Discussion}
Effect of co-doping Ni with Co on \textit{D} and lattice of the Sn$ _{0.97-\text{y}} $Co$ _{0.03} $Ni$ _{\text{y}} $O$ _{2} $ nanocrystals is clearly observed from Fig. \ref{figsize} and \ref{figlattice} respectively. The exponential decrease in \textit{D} (Fig. \ref{figsize}) with increase in \textit{y} can be attributed to growth kinetics involved during the nucleation process of the nanocrystals \cite{Roy}.  The decrease in \textit{D} is such that after \textit{y} = 0.02, \textit{D} approaches strong confinement regime (Bohr exciton radius for SnO$ _{2} \approx$ 2.5 nm) from weak confinement regime, marked by a strong-confinement signature of blue shift in E$ _{\text{g}}^{\text{allw.}} $ after \textit{y} = 0.02 (inset (b) of Fig. \ref{figreflectance}). To have an insight into the effect of co-doping on the lattice of the nanocrystals, Rietveld refinement of the XRD patterns for the nanocrystals (Fig. \ref{figxrd}) can be considered. From Fig. \ref{figxrd}, it can be observed that the Sn$ _{0.97-y} $Co$ _{0.03} $Ni$ _{y} $O$ _{2} $ nanocrystals retain the P4$ _{2} $/\textit{mnm} space group of tetragonal SnO$_{2}$ throughout the doping concentration (\textit{y}). In an ideal tetragonal SnO$_{2}$, Sn$^{4+} $ at 2a lattice site exists in distorted octahedral coordination geometry of O$^{2-}$ at 4f position and each O$^{2-}$ is in turn having a trigonal planar coordination geometry of Sn$^{4+} $ \cite{Kilic}. When, Co and Ni are doped at cationic site in SnO$  _{2}$, they can have either +2 or +3 as a possible charge state \cite{Roy,Srinivas}. The ionic radii of Co in octahedral geometry (assuming in low spin configuration) are 0.65 \AA (for Co$ ^{2+} $) and 0.545 \AA (for Co$ ^{3+} $); and that for Ni are as 0.69 \AA (for Ni$ ^{2+} $) and 0.56 \AA (for Ni$ ^{3+} $), whereas the ionic radius of Sn$ ^{4+} $ is 0.69 \AA \cite{Shannon}. Hence, when Co and Ni are co-doped at 2a lattice site of Sn in SnO$_{2}$, to maintain the tetragonal crystal symmetry, as in the present case, it is expected that, there will be an excess of positive charge at 2a site w.r.t. the original lattice (i.e., charge neutrality criterion would tend to be violated), as well as a tensile strain would be developed in the lattice, manifested by an increase in interplanar spacing (d$ _{hkl} $). As such, for the present Sn$ _{0.97-\text{y}} $Co$ _{0.03} $Ni$ _{\text{y}} $O$ _{2} $ nanocrystals, the larger ionic radius of the two types of dopants combined has led to an increase of the lattice parameter \textit{a} (= \textit{b}) (Fig.\ref{figlattice}(a)) and the corresponding increase in the unit cell volume (\textit{V}) (Fig.\ref{figlattice}(c)) is reflected in an enhancement of the average tensile lattice strain (inset (a) of Fig.\ref{figsize}), with an increase in d$ _{hkl} $, for example, d$ _{110} $ as shown in inset (b) of  Fig. \ref{figsize}. The variation of lattice parameters \textit{a} (= \textit{b}) and \textit{c} deviates from Vegard's law, which is expected since for confirming to Vegard's law, the host and the dopant need to have negligible disparity in the ionic sizes and electrochemical differences \cite{Denton}, which is not valid in the present nanocrystals. However, an opposite behaviour of \textit{c} (Fig.\ref{figlattice}(b)), with Ni doping concentration, as compared to \textit{a} (= \textit{b}) (Fig.\ref{figlattice}(a)) is observed and require an explanation from the geometry of point defects in the lattice of the nanocrystals.
%----------------------------------------------------------------------------------
\subsection{Identification of point defects}
%---------------------------------
\begin{figure}[ht]
\centering
\includegraphics[width=1.00\linewidth]{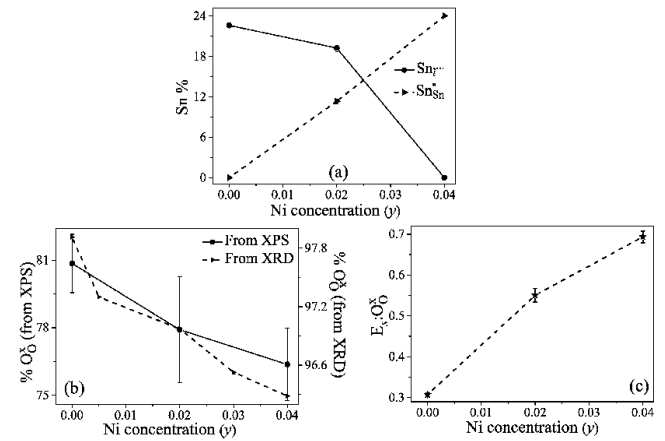}
\caption{Variation of \% of Sn$_{i}^{\centerdot\centerdot\centerdot\centerdot}$ and Sn$_{\text{Sn}}^{''}$ in total Sn concentration (a); \% of lattice oxygen (O$ _{\text{O}}^{\text{x}} $) as obtained from O1s core level XPS and Rietveld refinement of XRD (b); and ratio of concentration of surface state to that of lattice oxygen (E$ _{s} $ : O$ _{\text{O}}^{\text{x}} $) (c) with Ni doping concentration (\textit{y}). All the solid and dotted lines are eye guides.}
\label{figatconc}
\end{figure}
%-----------------------------------
 First principles density functional theory calculations for SnO$ _{2} $ have indicated that non-stoichiometric defects are of lower formation energies than stoichiometric defects and hence would dominate the resulting defect structure\cite{Godinho}. Thus, for the present nanocrystals, the presence of non-stoichiometric defects can be expected.\\\\ 
\textbf{\textit{Non-stoichiometric defect}:} As stated above, in an ideal crystal lattice of SnO$ _{2} $, Sn$ ^{4+} $ exists at 2a lattice sites with distorted octahedral coordination geometry of O$ ^{2-} $, situated at 4f sites. These Sn$ ^{4+} $ at 2a sites and O$ ^{2-} $ at 4f sites, being charge neutral w.r.t. the original lattice can be designated as Sn$ _{\text{Sn}}^{\text{x}} $ and O$ _{\text{O}}^{\text{x}} $ respectively, in accordance with Kroger-Vink notations \cite{Kroger}. The existence of Sn$ _{\text{Sn}}^{\text{x}} $ in the present Sn$ _{0.97-\text{y}} $Co$ _{0.03} $Ni$ _{\text{y}} $O$ _{2} $ nanocrystals is evident from the occurrence of the Sn 3d$ _{5/2} $ core level peak at $ \approx $ 486.342 eV and the occurrence of O1s core-level peak at $ \approx $ 529.273 eV evident the presence of O$ _{\text{O}}^{\text{x}} $ in the nanocrystals. As can be seen from Fig. \ref{figsn3d}, for Ni0 and Ni2 nanocrystals, a Sn 3d$ _{5/2} $ core level peak corresponding to metallic Sn has appeared on the lower binding energy side to Sn$ _{\text{Sn}}^{\text{x}} $ at around 485.034 eV for Ni0 and 484.981 eV for Ni2 nanocrystals. Since, this peak appears in addition to that corresponding to Sn$ _{\text{Sn}}^{\text{x}} $, it can be Sn going to an interstitial site, i.e., Sn$ _{i} $. Now, charge neutral Sn$ _{i} $ requires a much higher formation energy than Sn$ _{i} $ in +4 charge state for to be placed in SnO$ _{2} $ lattice \cite{Kilic}. Hence, the corresponding point defect can be identified as Sn$_{i}^{\centerdot\centerdot\centerdot\centerdot}$ -- a non-stoichiometric interstitial Sn defect, which has been predicted based on first-principles calculations to be a stable defect in SnO$ _{2} $ \cite{Kilic} and has also been observed earlier for Co-doped SnO$ _{2} $ nanocrystals \cite{Roy}. However, the point defect Sn$_{i}^{\centerdot\centerdot\centerdot\centerdot}$ formed in the lattice cannot address the charge imbalance occurring on co-doping in the present nanocrystals and requires lowering of some positive charge at the sites of the host cation. Thus, to fulfill the charge neutrality condition, there occurs a partial reduction of Sn$ _{\text{Sn}}^{\text{x}} $ to Sn$ _{\text{Sn}}^{''} $, as evident from the occurrence of the 3d peak corresponding to SnO in the Sn3d core-level spectra of Ni2 and Ni4 nanocrystals (Fig. \ref{figsn3d}) as well as the existence of local structures resembling to SnO in HR-TEM image of the Ni3 nanocrystals (Fig. \ref{figtem}(c)). Although the Ni co-doping do not affect appreciably the core-level electronic configurations of all the three forms of Sn (Sn$ _{\text{Sn}}^{\text{x}} $, Sn$_{i}^{\centerdot\centerdot\centerdot\centerdot}$ and Sn$ _{\text{Sn}}^{''} $), as indicated by the absence of any significant chemical shift in the Sn3d core level spectra (Fig. \ref{figsn3d}) of the nanocrystals, but interestingly, concentration of the point defect Sn$ _{\text{Sn}}^{''} $ has been found to increase in compensation of Sn$_{i}^{\centerdot\centerdot\centerdot\centerdot}$ with increase in Ni co-doping concentration (\textit{y}), as shown in Fig. \ref{figatconc}(a). The increment is such that for Ni0 nanocrystals, only Sn$_{i}^{\centerdot\centerdot\centerdot\centerdot}$ exist; for Ni2 nanocrystals both the point defects -- Sn$ _{\text{Sn}}^{''} $ and Sn$_{i}^{\centerdot\centerdot\centerdot\centerdot}$ coexist in the lattice whereas with increase in Ni concentration for Ni4 nanocrystals, only Sn$ _{\text{Sn}}^{''} $ remains along-with Sn$ _{\text{Sn}}^{\text{x}} $ in the nanocrystals (Figs. \ref{figsn3d} and \ref{figatconc}(a)). This emphasizes the importance of maintaining charge neutrality in the lattice of the nanocrystals on increasing the Ni co-doping concentration. However, the combined effect of localised nature of these point defects in the lattice of the nanocrystals, the very small ratio of the defects Sn$_{i}^{\centerdot\centerdot\centerdot\centerdot}$ and Sn$ _{\text{Sn}}^{''} $ to Sn$ _{\text{Sn}}^{\text{x}} $ and the existence of local disorder of SnO in small average crystallite size of the nanocrystals makes the SnO-resembling local structures to be directly non-observable in the diffraction experiments -- XRD (Fig. \ref{figxrd}) and SAED (Fig. \ref{figtem}(b)) for the nanocrystals, but their effect, in conjunction with other point defects can be observed, as discussed later.\\\\ 
\textbf{\textit{Stoichiometric anion Frenkel defect}:} In addition to the non-stoichiometric Sn related point defects -- Sn$_{i}^{\centerdot\centerdot\centerdot\centerdot}$ and Sn$ _{\text{Sn}}^{''} $, there also exists O related Frenkel-type stoichiometric point defects, i.e., oxygen-vacancies (\textit{V}$ _{\text{O}} $) and oxygen-interstitials (O$ _{i} $). In fact, \textit{V}$ _{\text{O}} $ is considered to be a very stable intrinsic point defect in SnO$ _{2} $, such that SnO$ _{2} $ is usually expressed in its non-stoichiometric form SnO$ _{2-\delta} $ \cite{Kilic, Godinho}. The presence of \textit{V}$ _{\text{O}} $ in the lattice of the nanocrystals can be observed from the O1s core level spectra (Fig. \ref{figo1s}) as well as Rietveld refinement of the XRD patterns (Fig. \ref{figxrd}) for the nanocrystals. Figure \ref{figatconc}(b) shows the qualitative variation of \textit{V}$ _{\text{O}} $ with increase in Ni co-doping concentration (\textit{y}). The concentration of  \textit{V}$ _{\text{O}} $ qualitatively increases  with increase in Ni co-doping concentration (\textit{y}). Now \textit{V}$ _{\text{O}} $ can have three charge states (charge w.r.t. the original lattice) -- the neutral \textit{V}$ _{\text{O}} $ (\textit{V}$ _{\text{O}}^{\text{x}} $), singly-ionised \textit{V}$ _{\text{O}} $ (\textit{V}$ _{\text{O}} ^{\centerdot}$) and doubly-ionised \textit{V}$ _{\text{O}} $ (\textit{V}$ _{\text{O}}^{\centerdot\centerdot}$). Among them, only \textit{V}$ _{\text{O}} ^{\centerdot}$ pose as paramagnetic centres \cite{Soumen} and hence, contribute to the paramagnetism in the nanocrystals. In fact, \textit{V}$ _{\text{O}} ^{\centerdot}$ are considered to enhance the ferromagnetic interactions in diluted magnetic oxide-semiconductor nanocrystals, as explained using bound magnetic polaron model \cite{Brijmohan2}. From the inset of Fig. \ref{figmagnetic}, Curie-Weiss fit predicts an antiferromagnetic interaction below 300 K for Ni0 nanocrystals whereas a ferromagnetic interaction for Ni3 nanocrystals. Hence, from this changeover of antiferromagnetic interactions to ferromagnetic interactions, it can be concluded that on increasing the  Ni co-doping concentration (\textit{y}), concentration of \textit{V}$ _{\text{O}} ^{\centerdot}$ increases in the nanocrystals and that they form the dominant point defect among all the charge states of \textit{V}$ _{\text{O}} $. Regarding the anion Frenkel counterpart of the \textit{V}$ _{\text{O}} $, i.e., the oxygen-interstitials (O$ _{i} $), they are considered to relax to form dangling bonds as peroxide ion (O$_{2}^{2-}$ and O$ ^{-} $) type structures\cite{Godinho}, which move to the surface of the nanocrystals to form surface defect states (E$ _{s} $)\cite{Patra3}. The existence of such extra non-lattice oxygen is evident from the occurrence of the peak at higher binding energy to the peak for O$ _{\text{O}}^{\text{x}} $ for the nanocrystals (Fig. \ref{figo1s}). Interestingly, the concentration of  E$ _{s} $ has been found to increase with increase in Ni co-doping concentration (\textit{y}), as can be observed from the qualitative increment in the ratio of  E$ _{s} $ to O$ _{\text{O}}^{\text{x}} $ with \textit{y} in Fig. \ref{figatconc}(c). The simultaneous increment (qualitatively) of \textit{V}$_{\text{O}}$ and E$ _{s} $ (Figs. \ref{figatconc}(b) \& (c) resp.) clearly establishes the anion Frenkel-type nature of these oxygen related point defects. In fact, a similar trend has also been observed earlier for Co doped SnO$ _{2} $ nanocrystals \cite{Roy} and has been extensively studied for Cu doped SnO$ _{2} $ based memristors \cite{Meifang}.        

All these point defects (Sn related and O related) have both microscopic and macroscopic influence on the Sn$ _{0.97-\text{y}} $Co$ _{0.03} $Ni$ _{\text{y}} $O$ _{2} $ nanocrystals.
%-----------------------------------------------------------------------------------------
\subsection{Microscopic influence of the point defects}
%--------------------------------
\begin{figure}[ht]
\centering
\includegraphics[width=1.00\linewidth]{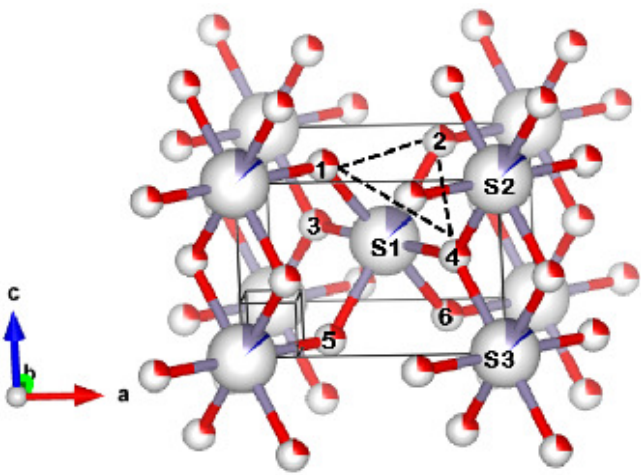}
\caption{The VESTA derived structure (using ball-and-stick model) for Ni3 nanocrystals. The atoms \textit{viz.} 1, 2, 3, 4, 5, 6 denote the O$ _{\text{O}}^{\text{x}} $ and atom \textit{viz.} S denote Sn$ _{\text{Sn}}^{\text{x}} $ lattice position. Atoms \textit{viz.} 1, 2, 3, 4, 5, 6 form the 6 vertices of the  central Sn \textbf{--} O octahedron. The dotted lines form $ \vartriangle $124 as one of the 8 triangular faces of the octahedron.}
\label{figvesta}
\end{figure}
%---------------------------------
The microscopic influence of the point defects is concerned with the effect on geometry of Sn -- O octahedron as well as on the trigonal planar coordination geometry of O in the nanocrystals. For the octahedral coordination geometry of Sn, it has been found that with Ni co-doping, although the distortion of the complete Sn -- O octahedron reduces, but, the Sn -- O bonds of the octahedron tends to form a distorted tetrahedral geometry resembling SnO. However, the overall distortion in the trigonal planar coordination geometry of O has been found to increase with increase in Ni co-doping concentration (\textit{y}). This can be understood considering the effect of point defects on the bond parameters, i.e., bond lengths and bond angles. From Fig.\ref{figvesta}, it can be seen that in the slightly distorted Sn \textbf{--} O octahedron w.r.t. the central atom Sn$ _{\text{Sn}}^{\text{x}} $ at (0.5, 0.5, 0.5) (\textit{viz.} atom S1), four O$_{\text{O}}^{\text{x}}$, \textit{viz.} atoms 1, 2, 5 and 6 lying on \textit{ab}-plane are at the equatorial positions, each shared by two unit cells and two O$_{\text{O}}^{\text{x}}$, \textit{viz.} atoms 3 and 4 lying completely inside the unit cell in a plane parallel to \textit{c}-axis are at axial positions. For considering the trigonal planar coordination geometry of O, a typical O$_{\text{O}}^{\text{x}}$ in the Fig. \ref{figvesta}, say, \textit{viz.} atom 4, can be chosen, where the trigonal planar coordination geometry is formed along-with  Sn$ _{\text{Sn}}^{\text{x}} $ at positions, \textit{viz.} S1, S2 and S3.\\\\
%---------------------------
\begin{figure}[hb]
\centering
\includegraphics[width=1.00\linewidth]{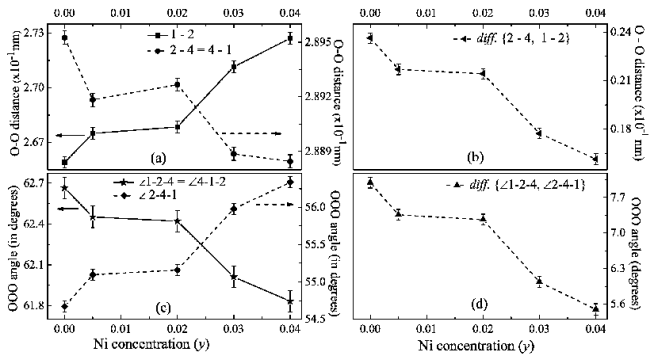}
\caption{The variation of various O \textbf{--} O bond parameters with Ni co-doping concentration (\textit{y}). The syntax \textit{diff} $\lbrace  $\textit{quantity1},\textit{quantity2}$\rbrace$  stands for the difference between \textit{quantity1} and \textit{quantity2}. The solid and dotted lines are eye guides.}
\label{figoo}
\end{figure}
%---------------------------
\begin{figure}[ht]
\centering
\includegraphics[width=1.00\linewidth]{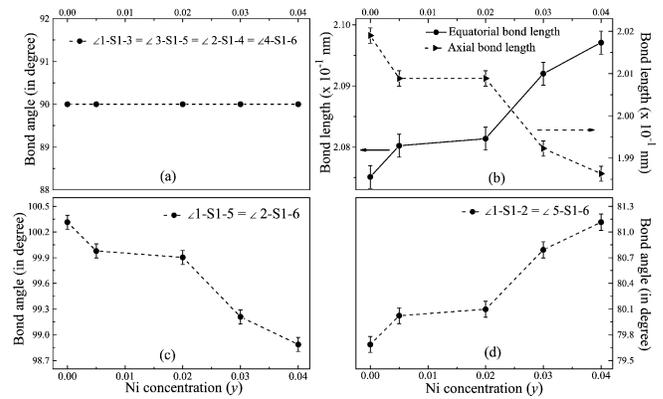}
\caption{The variation of various Sn \textbf{--} O bond parameters, in Sn octahedral coordination geometry, with Ni co-doping concentration (\textit{y}). The solid and dotted lines are eye guides.}
\label{figosn}
\end{figure}
%---------------------------
\textbf{\textit{Effect on octahedral coordination geometry of Sn}:} Now, by geometry, a regular octahedron is built up of six equilateral triangles. When distortion in the octahedron occurs, then a distortion in each of the equilateral triangle can be expected. A such typical triangle can be considered in Fig.\ref{figvesta}, shown by dotted lines, formed by the O$ _{\text{O}}^{\text{x}} $ atoms \textit{viz.} 1, 2 and 4 at the vertices. For the triangle to be an equilateral triangle, distance 1 \textbf{--} 2 = 2 \textbf{--} 4 = 4 \textbf{--} 1; and $ \angle $1-2-4 = 60$^\circ$ = $ \angle $2-4-1 = $ \angle $4-1-2. Figure \ref{figoo}(a) and (c) shows the variation of typical edge lengths and angles for the triangle with \textit{y}. It can be seen from Fig.\ref{figoo}(a) that 2 \textbf{--} 4 = 4 \textbf{--} 1 $ \neq $ 1 \textbf{--} 2; however the difference (Fig.\ref{figoo}(b)) decreases with increase in \textit{y}. Similarly, $ \angle $1-2-4 = $ \angle $4-1-2 $ \neq $ $ \angle $2-4-1; the difference between the angles (Fig. \ref{figoo}(d)) being decreasing with increase in \textit{y}. Thus, the isosceles triangle $ \bigtriangleup $1-2-4 tends to an equilateral triangle with increase in \textit{y}. Hence, it can be concluded that Ni co-doping leads to a decrease in the distortion of the complete Sn \textbf{--} O octahedron, where all the Sn \textbf{--} O bonds and the point defects \textbf{--} V$ _{\text{O}}^{\centerdot}$ and Sn$ _{\text{Sn}}^{''} $ are included. For the geometry formed by only Sn \textbf{--} O bonds in the octahedron on co-doping Ni, a prediction can be made based on the variation of bond angle $\angle$1-S1-5, formed between O$ _{\text{O}}^{\text{x}} $ at 1 and 5 to Sn$ _{\text{Sn}} $ at S1 in Fig. \ref{figvesta}. For that the position of V$ _{\text{O}}^{\centerdot}$ need to be determined. It can be seen from Fig. \ref{figvesta} that the point defect V$ _{\text{O}}^{\centerdot}$ can be created at the equatorial or axial position. This position plays a crucial role in defining the geometry of the Sn \textbf{--} O bonds as well as the variation of \textit{c}-axis with \textit{y}. It can be observed from Fig. \ref{figosn}(a) that $ \angle $1-S1-3, $ \angle $3-S1-5 always remain at 90$^\circ $ throughout the co-doping concentration. Similar is the case for $ \angle $2-S1-4 and $ \angle $4-S1-6. This would not have been, if V$ _{\text{O}}^{\centerdot}$ is created at the equatorial position, at say, position 1 or 2, because in that case the only repulsion between O$ _{\text{O}}^{\text{x}} $ at positions \textit{viz.} 3 and 5 or between 4 and 6 would have increased the bond angle greated than 90$^\circ $. Thus, it can be assumed that V$ _{\text{O}}^{\centerdot}$ is created at the axial position in the octahedron. Due to the V$ _{\text{O}}^{\centerdot}$ created at an axial position, the size of the lattice site w.r.t. the site having O$ _{\text{O}}^{\text{x}} $ in original lattice reduces, hence the axial bond length must decrease, as shown in Fig. \ref{figosn}(b). Now, the co-dopants can be substituted at a cationic lattice site, i.e., either the central Sn$ _{\text{Sn}}^{\text{x}} $ atom at (0.5,0.5,0.5) or any of the corner Sn atoms, say, \textit{viz.} atom S2 or S3 (Fig. \ref{figvesta}). Similarly, the point defect Sn$ _{\text{Sn}}^{''} $ would be created at at cationic lattice site \textit{viz.} S1, or site \textit{viz.} S2 or S3. Since, the combined ionic radius of the dopants or the size of Sn$ _{\text{Sn}}^{''} $  is larger than that of Sn$ _{\text{Sn}}^{\text{x}} $, hence under both the substitutions, the overall size of a cationic lattice site (\textit{viz.} S1 or S2 or S3 in Fig. \ref{figvesta}) increases. Considering the cationic lattice site, \textit{viz.} S1 in Fig. \ref{figvesta}, when its size increases, the intersite distance between sites, \textit{viz.} S1 with 1, S1 with 2, S1 with 5 or S1 with 6 effectively increases. This in combination with the creation of V$ _{\text{O}}^{\centerdot}$ at an axial position, which leads to an outward relaxation of the three nearest neighbour Sn$ _{\text{Sn}}^{\text{x}} $ \cite{AKSingh} eventually increases the equatorial bond length (Fig. \ref{figosn}(b)). As such, the electron density on the central cationic site decreases, i.e., the polarisation of the bond decreases, thereby reducing the covalent character of the Sn -- O bond \cite{Fajan}. Thus in a combined effect of a reduction in the electron density on the cationic site  as well as reduction in the net effective repulsion between the equatorial oxygen atom pairs at \textit{viz.} 1 with 5 and 2 with 6 as compared to that  between  the equatorial oxygen atom pairs at \textit{viz.} 1 with 2 and 5 with 6 on V$ _{\text{O}}^{\centerdot}$ being created at an axial position, the bond angles $ \angle $1-S1-5 = $ \angle $2-S1-6 effectively decreases (Fig. \ref{figosn}(c)) whereas the bond angles $ \angle $1-S1-2 = $ \angle $5-S1-6 (Fig. \ref{figosn}(d)) effectively increases. As such, the equatorial oxygen atom pairs at \textit{viz.} 1 and 5, or, 2 and 6 might leave the \textit{ab}-plane to come closer to each other for a more compact and stable geometry. Thus, the resultant geometry resembles to a distorted tetrahedron, like SnO \cite{Kilic}. Hence, it can be concluded that the creation of point defects V$ _{\text{O}}^{\centerdot}$ at axial position and Sn$ _{\text{Sn}}^{''} $  on Ni co-doping affects the distorted Sn \textbf{--} O octahedral geometry in a way that the octahedral distortion reduces. However, if the geometry formed by the Sn \textbf{--} O bonds is considered, then they seem to form a distorted tetrahedral geometry, mimicing SnO, from the existing distorted octahedral geometry of SnO$ _{2} $. The formation of local SnO-resembling structures is also confirmed from HR-TEM, where lattice spacing corresponding to (111) and (112) lattice planes of orthorhombic SnO phase have been observed, along-with (110) planes of tetragonal SnO$ _{2} $ phase (Fig. \ref{figtem}(c)). The stable multi-valence nature of Sn supports this coexistence of SnO and SnO$ _{2} $ phases in the nanocrystals \cite{Kilic, Godinho}.\\\\
\textbf{\textit{Effect on trigonal planar coordination geometry of O}:} 
%---------------------------------------------
\begin{figure}[h]
\centering
\includegraphics[width=1.00\linewidth]{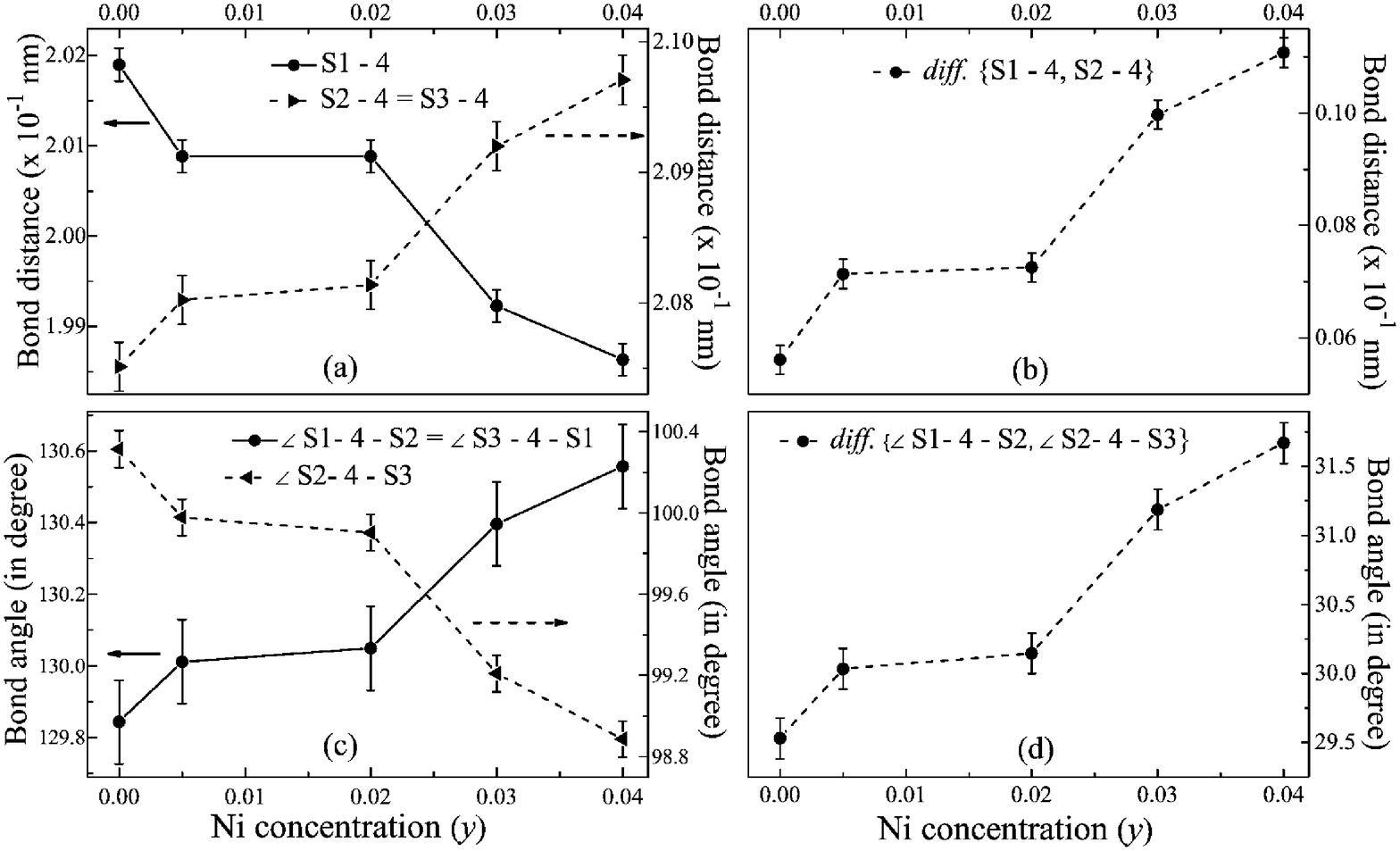}
\caption{The variation of various Sn \textbf{--} O bond parameters, in the trigonal planar coordination geometry of O, with Ni co-doping concentration (\textit{y}). The syntax \textit{diff} $\lbrace  $\textit{quantity1},\textit{quantity2}$\rbrace$  stands for the difference between \textit{quantity1} and \textit{quantity2}. The solid and dotted lines are eye guides.}
\label{figotriangle}
\end{figure}
%----------------------------------------------
By geometry, a trigonal planar coordination geometry is formed by four atoms -- a central atom and three similar atoms at the periphery, all four lying on the same plane, with all the three bond lengths being equal and all the bond angles being equal to 120$^\circ $. Figure \ref{figotriangle} shows the variation of the bond lengths and angles for the trigonal planar coordination geometry of a typical O$ _{\text{O}}^{\text{x}} $ at position, \textit{viz.} atom 4 in Fig. \ref{figvesta}. It can be seen that the bond length S2 -- 4 = S3 -- 4 $ \neq $ S1 -- 4 (Fig. \ref{figotriangle}(a)); and the bond angle $ \angle $S1-4-S2 = $ \angle $S3-4-S1 $ \neq \angle $S2-4-S3 (Fig. \ref{figotriangle}(c)). Thus, there lies a distortion in the trigonal planar coordination geometry. With Ni co-doping, since the overall electronic polarisation of bond decreases, hence, the bond angle $\angle $S2-4-S3 decreases, with an increase in $ \angle $S1-4-S2 = $ \angle $S3-4-S1. The change in bond lengths and bond angles occurs in a way that the difference between the bond lengths S2 -- 4 (= S3 -- 4) and S1 -- 4 as well as that between bond lengths $ \angle $S1-4-S2 (= $ \angle $S3-4-S1) and $ \angle $S2-4-S3 increases with increase in Ni co-doping concentration (\textit{y}) (Figs. \ref{figotriangle}(b) and (d) resp.). Hence, it can be concluded that Co-Ni co-doping leads to an enhancement of the distortion of trigonal planar coordination geometry of O in SnO$ _{2} $ nanocrystals.
%--------------------------------------------------------------------------------
\subsection{Macroscopic influence of the point defects}
The macroscopic effect of point defects on structural properties of the nanocrystals is observed from Figs. \ref{figlattice} and \ref{figurbach}(f). Due to changes in the geometry of Sn \textbf{--} O octahedron and the trigonal planar coordination geometry of O, overall distortion in the lattice increases, as evident from the increase in Urbach energy (E$ _{u} $) \cite{Roy, Sajeev} with increase in \textit{y}(Fig.\ref{figurbach}(f)). As can be seen from Fig. \ref{figosn}(d), the equatorial Sn \textbf{--} O bond length increases by $ \approx $ (1.063$ \pm $0.009)\% with increase in \textit{y} (Fig.\ref{figosn}(d)). However, the resultant increase in the equatorial bond length is small as compared to the increase in $ \angle $1-S-2 = $ \angle $5-S-6 ($ \approx $ (1.811$ \pm $0.041\%) (Fig. \ref{figosn}(a)) with increase in \textit{y}. This causes a decrease in \textit{c}-axis, as observed in Fig. \ref{figlattice}(b). However, the overall effect of larger ionic radius at the cationic site increases the lattice parameter \textit{a} (= \textit{b}). 

A profound effect of the point defects on emission properties of the nanocrystals can be observed from Figs. \ref{figpl} and \ref{figspectral}. The UV-emission peak in the room temperature PL spectra of the nanocrystals (Fig. \ref{figpl}) indicates the breaking of the dipole forbidden selection rule \cite{Roy}. Existence of all the emissions below E$ _{\text{g}}^{\text{allw.}} $ of the nanocrystals suggests inclusion of defect trap states in the sub-band gap region, which act as trapping centres for the electrons relaxing from conduction band. It has been proposed that among the possible point defects in SnO$ _{2} $, only \textit{V}$ _{\text{O}} $ forms energy levels below the conduction band minimum (CBM) \cite{Kilic}. Now, as discussed above, for the present nanocrystals, \textit{V}$ _{\text{O}} $ has been found to exist as \textit{V}$ _{\text{O}} ^{\centerdot}$, the concentration of which has been found to increase with increase in \textit{y} (Fig. \ref{figatconc}(b)). Also, the concentration of surface defect states, E$ _{s} $, which form the Frenkel counterpart of \textit{V}$ _{\text{O}} ^{\centerdot}$, i.e., O$ _{i} $ has been found to increase with Ni co-doping concentration (\textit{y}), as shown in Fig. \ref{figatconc}(c). Kar, \textit{et al.} \cite{Patra3} and Gu, \textit{et al.} \cite{F_Gu} proposed the role of E$ _{s} $ along-with V$ _{\text{O}}^{\centerdot}$ centres for the origin of blue luminescence in SnO$ _{2} $ nanocrystals, whereas, Zhou \textit{et al.} \cite{XT_Zhou} proposed the transition of electrons from CBM to E$ _{s} $ for the observed blue emission in the SnO$ _{2} $ nanocrystals. The origin of blue luminescence for the present Sn$ _{0.97-\text{y}} $Co$ _{0.03} $Ni$ _{\text{y}} $O$ _{2} $ nanocrystals can be due to a combination of both the mechanisms proposed above, following the schematic energy level diagram in ref. \cite{Roy}. A direct effect of the increase in concentration of point defects in the nanocrystals is thus an enhancement of the spectral contribution of blue luminescence by $ \approx $ 72.33\%, as shown in Fig. \ref{figspectral}(a). The green luminescence involves the transition of electrons from \textit{V}$ _{\text{O}}^{\text{x}} $ to E$ _{s} $ and  \textit{V}$ _{\text{O}}^{\centerdot\centerdot}$/ \textit{V}$ _{\text{O}} ^{\centerdot}$ \cite{Roy}. Since, as discussed, for the present nanocrystals, there exists a much high concentration of \textit{V}$ _{\text{O}} ^{\centerdot}$ at the expense of other charge states of \textit{V}$ _{\text{O}} $, hence the spectral contribution of green emission decreases by $ \approx $ 40.49\% with Ni co-doping concentration (\textit{y}) (Fig.\ref{figspectral}(b)). However, the decrement of spectral contribution of green emissions need to be investigated further.  

%------------------------------------------------------------------------------------------------------------------------------------------------------------------------------------------------------------

\section{Summary \& Conclusions}
Cobalt, nickel co-doped SnO$ _{2} $ nanocrystals (Sn$ _{0.97-\text{y}} $Co$ _{0.03} $Ni$ _{\text{y}} $O$ _{2} $; 0.00 $ \leq $ \textit{y} $ \leq $ 0.04) have been successfully prepared via the chemical co-precipitation technique. Both non-stoichiometric and anion Frenkel-type stoichiometric point defects have been found to exist in the nanocrystals for only Co doped nanocrystals (\textit{i.e.}, \textit{y} = 0.00) as well as on co-doping with Ni (\textit{i.e.}, for  \textit{y}$ > $ 0.00). However, unlike for only Co doping, on increasing the Ni co-doping concentration, the already existing non-stoichiometric defect of Sn$ _{i}^{\centerdot\centerdot\centerdot\centerdot} $ gets systematically converted to Sn$_{Sn} ^{''} $, which has been confirmed from the SnO-related peak in Sn3d core-level XPS spectra and the formation of some local structures resembling SnO in the overall SnO$ _{2} $ lattice of the nanocrystals in HR-TEM image of the nanocrystals. Regarding the anion Frenkel defect, from the changeover of existing antiferromagnetic interactions below 300 K in \textit{y} = 0.00 to ferromagnetic interactions with increase in Ni co-doping concentration, the concentration of  \textit{V}$ _{\text{O}}^{\centerdot}$ has been concluded to be increasing at the compensation of other charge states of \textit{V}$ _{\text{O}}$ with increase in \textit{y}. The anion Frenkel counterpart of \textit{V}$ _{\text{O}}^{\centerdot}$, manifested in the form of surface defect states, E$ _{s} $, has also been found to increase with Ni co-doping. The combined effect of the non-stoichiometric and stoichiometric point defects results a reduction in the distortion of the overall Sn -- O octahedral geometry of SnO$ _{2} $ in a way that the Sn -- O bonds tend to form a distorted Sn -- O tetrahedron, mimicing SnO, whereas the distortion in trigonal planar coordination geometry of O increases with \textit{y}. As a macroscopic influence of the point defects on the nanocrystals, the lattice parameter \textit{c} has been found to decrease slightly with Ni co-doping as well the disorder in the lattice of the nanocrystals increases with \textit{y}, marked by an increase in the dominance of Urbach type absorption. Also, due to the anion Frenkel defect, the spectral contribution of blue luminescence in the total emission intensity has been found to increase by $ \approx $ 72\% for \textit{y} = 0.04 as compared to \textit{y} = 0.00.

%------------------------------------------------------------------------------------------------------------------------------------------------------------------------------------------------------
\begin{acknowledgments}
SR acknowledges CSIR, India for the Senior Research Fellowship vide file no. 09/013(0849)/2018 -- EMR-I. AKG is thankful to DST-FIST program; to DST-PURSE program; to UGC-UPE program; to UGC-CAS program; AKG is also thankful to DAE-BRNS, India; CSIR, India; and UGC, India for financial support (Grant no. 2011/37P/11/BRNS/1038-1, 03(1302)/13/EMR-II, and F: 42-787/2013 (SR) respectively), to the Bio-Physics lab, Deptt. of Physics (B.H.U.) for PL facility, to the Laboratory for Central Facilities, Deptt. of Physics (B.H.U.) for VSM based room temperature magnetisation measurements and to the Central Instrument Facility, IIT-BHU for SQUID based temperature dependent magnetisation measurements.
\end{acknowledgments}

%-----------------------------------------------------------------------------------------------------------------------------------------------------------------------------------------------------------

%\nocite{*}
\bibliographystyle{aipnum4-1}
\bibliography{ref_jap}

\end{document}